\documentclass[journal = nalefd, manuscript=letter, layout=twocolumn,email = true]{achemso}

\usepackage[version=3]{mhchem} 
\usepackage[T1]{fontenc}       
\usepackage{color}     

\author{Taiki Yoda}
\author{Takehito Yokoyama}
\author{Shuichi Murakami}
\email{murakami@stat.phys.titech.ac.jp}
\affiliation[Department of Physics, Tokyo Institute of Technology]{Department of Physics, Tokyo Institute of Technology, Ookayama, Meguro-ku, Tokyo 152-8551, Japan}
\alsoaffiliation[TIES, Tokyo Institute of Technology]{TIES, Tokyo Institute of Technology, Ookayama, Meguro-ku, Tokyo 152-8551, Japan}

\title[OEE]
{Orbital Edelstein effect as a condensed-matter analog of solenoid}


\begin{document}




\begin{abstract}
We theoretically study current-induced orbital magnetization in a chiral crystal.
This phenomenon is an orbital version of the Edelstein effect.
We propose an analogy between the current-induced orbital magnetization and an Amp\`ere field in a solenoid in classical electrodynamics.
In order to quantify this effect, we define a dimensionless parameter from the response coefficients relating a current density with an orbital magnetization.
This dimensionless parameter can be regarded as a number of turns within a unit cell when the crystal is regarded as a solenoid, and it represents how ``chiral'' the crystal is.
By focusing on the dimensionless parameter, one can design band structure which realizes induction of large orbital magnetization.
In particular, a Weyl semimetal with all the Weyl nodes close to the Fermi energy can have a large value of this dimensionless parameter, which can exceed that of a classical solenoid.
\end{abstract}


Coupling between charge and spin degrees of freedom leads to various conversion phenomena between charge current and spin.
Typical examples of the conversion are the spin-Hall effect \cite{Murakami1348,PhysRevLett.92.126603,RevModPhys.87.1213,manchon2015new} and the Edelstein effect \cite{Edelstein90,Aronov91,Inoue03,RevModPhys.87.1213,manchon2015new}.
These effects make it possible to control magnetization by the charge current.
However, their magnitudes are limited by the size of the spin-orbit interaction since they are driven by the spin-orbit interaction.

Recently, a different mechanism of conversion between a charge current and a magnetization has been proposed. We proposed current-induced orbital magnetization \cite{yoda2015current} and Zhong et al. proposed gyrotropic magnetic effect \cite{PhysRevLett.116.077201}.
These effects are described by a similar response coefficient. 
In particular, the current-induced orbital magnetization is an orbital analog of the Edelstein effect, and 
we can call this effect an orbital Edelstein effect. 
In the orbital Edelstein effect, we focus on the orbital magnetic moment of the Bloch states given by \cite{PhysRevB.53.7010,PhysRevB.59.14915,RevModPhys.82.1959}
\begin{equation}
	{\bf m}_{n{\bf k}} = \frac{e}{2\hbar} \text{Im}
		\langle \partial_{{\bf k}} u_{n{\bf k}} |
		\times [ H_{{\bf k}} - \varepsilon_{n{\bf k}} ] |
		\partial_{{\bf k}} u_{n{\bf k}} \rangle,
	\label{Orbital_moment}
\end{equation}
where $H_{{\bf k}}$ is the Bloch Hamiltonian with eigenvalues $\varepsilon_{n{\bf k}}$, and $| u_{n{\bf k}} \rangle$ is the periodic part of a Bloch state in the $n$th band.
This orbital magnetic moment ${\bf m}_{n{\bf k}} $ is associated with each Bloch state $|u_{n{\bf k}}\rangle$. It is an expectation 
value of the operator of the 
orbital magnetic moment $-\frac{e}{2}{\bf r}\times{\bf v}$ taken for the 
Wannier orbital corresponding to the Bloch state, where $-e$ is 
an electron charge, ${\bf r}$ is the position of an electron and ${\bf v}$ is the velocity of an
electron \cite{PhysRevB.53.7010,PhysRevB.59.14915,RevModPhys.82.1959}. 
Hereafter, we assume that the time-reversal symmetry is preserved, which yields ${\bf m}_{n{\bf k}}=- {\bf m}_{n,-{\bf k}}$. 
If the inversion symmetry is also preserved, it yields ${\bf m}_{n{\bf k}}= {\bf m}_{n,-{\bf k}}$, leading to ${\bf m}_{n{\bf k}}\equiv 0$ for all the bands. Instead, we here assume
that the inversion symmetry is broken; it hereby leads to nonzero ${\bf m}_{n{\bf k}}$ in general. 
In particular, in chiral crystals as we show later, this nonzero ${\bf m}_{n{\bf k}}$
naturally follows from the chirality of the crystals. 
Thus, in equilibrium, the 
total orbital magnetization
 for the whole system is zero because of cancellations 
between the contributions from ${\bf k}$ and $-{\bf k}$.
The distribution of the orbital magnetization in $k$ space
is similar to the distribution of 
the spin polarization
in spin-split bands in systems with spin-orbit coupling, such as Rashba systems and 
surfaces of topological insulators. In such systems with the spin-split bands, 
a charge current 
induces an unbalance between the populations at ${\bf k}$ and at $-{\bf k}$, 
and the total spin polarization becomes nonzero, which is called the spin Edelstein effect. Likewise, 
from Eq.~(\ref{Orbital_moment}), the orbital Edelstein effect is expected in a similar way,
if we consider the distribution of the 
orbital magnetization ${\bf m}_{n{\bf k}}$ in $k$ space instead of that of the spin polarization.

At zero temperature, the orbital Edelstein effect is formulated as a Fermi-surface integral of the orbital magnetization \cite{yoda2015current,PhysRevLett.116.077201},
\begin{eqnarray}
	M_{i} &=& \alpha^{\text{ME}}_{ij} E_{j},
	\\
	\alpha^{\text{ME}}_{ij} &=& e \tau \sum_{n} \int_{\text{BZ}} \frac{d{\bf k}}{(2\pi)^{3}}
		\frac{ df } { d\varepsilon } \Big|_{\varepsilon = \varepsilon_{n{\bf k}}} m_{n{\bf k}, i} v_{n{\bf k}, j},
	\label{alpha_ME}
\end{eqnarray}
where ${\bf M}$ is the orbital magnetization, ${\bf E}$ is the electric field, $\tau$ is the relaxation time, $f$ is the Fermi distribution function, 
$df / d\varepsilon |_{\varepsilon = \varepsilon_{n{\bf k}}} = - \delta(\varepsilon_{n{\bf k}} - \varepsilon_{\text{F}})$, $\varepsilon_{\text{F}}$ is the Fermi energy,
and ${\bf v}_{n{\bf k}} = (1/\hbar) \partial \varepsilon_{n{\bf k}} / \partial {\bf k}$ is the electron velocity.
Here, we adopted a relaxation-time approximation that $\tau$ is constant.
The tensor $\alpha^{\text{ME}}$ describing this response is an axial tensor with rank 2.
The class of crystals whose symmetry allows nonzero rank-2 axial tensor is called gyrotropic. 
In particular, breaking of inversion symmetry is required for gyrotropic crystals, and among 21 point groups lacking inversion symmetry, only 18 point groups are gyrotropic, having nonzero 
orbital Edelstein effect and nonzero (spin) Edelstein effect.
However, the physical origin of the orbital Edelstein effect is different from the Edelstein effect.
While the spin-orbit interaction is essential in the Edelstein effect, 
the orbital Edelstein effect does not need the spin-orbit interaction \cite{PhysRevLett.116.077201}.
The size of orbital Edelstein effect is determined by the lattice structure and hopping amplitudes between the sites in chiral crystals.
In chiral crystals, several interesting phenomena have been revealed: an electric response in a magnetic field such as magnetochiral anisotropy \cite{PhysRevLett.87.236602,pop2014electrical,doi:10.1063/1.1523895,PhysRevLett.117.146603} and a magnetic response in an electric field such as current-induced optical activity \cite{vorob1979optical,Shalygin2012}.

In this letter, to quantify this effect, we introduce a dimensionless parameter $\xi$, which corresponds to the number of turns within a unit cell when the crystal is regarded as a solenoid, and we then show that in some cases this dimensionless constant is 
much enhanced compared to its classical value.
To define the dimensionless constant $\xi$, we first introduce a tensor $\beta^{\text{Mj}}$ describing the ratio between a magnetization and an electric current density ${\bf j}$,
\begin{eqnarray}
	M_{i} = \beta^{\text{Mj}}_{ij} j_{j}.
	\label{beta_mj}
\end{eqnarray}
instead of $\alpha^{\text{ME}}$. 
Since both ${\bf M}$ and ${\bf j}$ are proportional to $\tau {\bf E}$, $\beta^{\text{Mj}}$ is independent of the relaxation time $\tau$ and can be determined by band structure only. 
We theoretically show that $\beta^{\text{Mj}}$ is largely enhanced when systems are in the Weyl semimetal phase and all the Weyl points are close to the Fermi energy, 
through our 
calculations in a tight-binding model and in an effective Weyl Hamiltonian.
We then define the dimensionless parameter $\xi$, by  expressing 
the tensor $\beta^{\text{Mj}}$ 
as a product between $\xi$ and a scale factor 
given by the lattice constants. 
$\xi$ indicates a ratio between a longitudinal and a circulating components of the electric current, and it represents an efficiency of the orbital Edelstein effect as compared with a classical solenoid.
These results are useful for designing band structure with large orbital Edelstein effect.
\begin{figure}
\includegraphics[width=8.46cm]{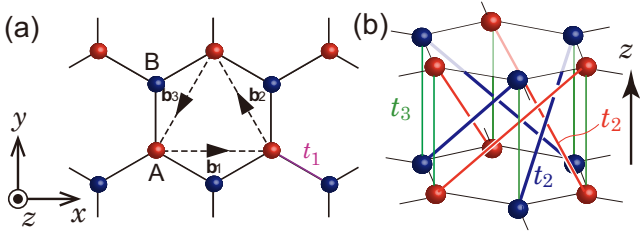}%
\caption{(a) One layer of the model forming a honeycomb-lattice.
Dashed arrows denote vectors ${\bf b}_{1}$, ${\bf b}_{2}$, and ${\bf b}_{3}$.
(b) Chiral hopping ($t_{2}$ term in Eq.~(\ref{TB_Hamiltonian})) in the right-handed helix.
Red (blue) lines denote hoppings between A (B) sites. }%
\label{fig_lattice}
\end{figure}

As an example, we here introduce a simple tight-binding model with chiral crystal structure proposed in Ref.~\citenum{yoda2015current}.
The tight-binding model is composed of honeycomb-lattice layers with one orbital per site, 
 as shown in Fig.~\ref{fig_lattice}(a), where ${\bf b}_{1} = a {\bf \hat{x}}$, ${\bf b}_{2} = a/2 (-{\bf \hat{x}} + \sqrt{3} {\bf \hat{y}})$, ${\bf b}_{3} = a/2 (-{\bf \hat{x}} - \sqrt{3} {\bf \hat{y}})$ and $a$ is a lattice constant in the honeycomb lattice.
The layers are stacked along the $z$-direction with an interlayer lattice constant $c$.
The Hamiltonian of the tight-binding model is written as
\begin{align}
	&H = t_{1} \sum_{\langle ij \rangle, l} c^{\dagger}_{i, l} c_{j, l}
		+ t_{3} \sum_{i, l} \sum_{s=\pm 1}c^{\dagger}_{i, l} c_{i, l+s}
	\nonumber \\
		&+ t_{2} \Biggl[ \sum_{i \in A, j, l} c^{\dagger}_{i + {\bf b}_{j}, l+1} c_{i, l} + \sum_{i \in B, j, l} c^{\dagger}_{i - {\bf b}_{j}, l+1} c_{i, l} 
		+ \text{H.c.}\Biggr],
	\label{TB_Hamiltonian}
\end{align}
where $c_{i,l}$ is an annihilation operator of an electron at the $i$th site in the $l$th layer, $t_{1}, t_{2},$ and $t_{3}$ are real constants, and we set $t_{1} > 0$ for simplicity.
The Hamiltonian (\ref{TB_Hamiltonian}) does not include spin-orbit interaction and spin indices are omitted.
The $t_{1}$ term is a nearest-neighbor hopping within the same honeycomb layers. The $t_{2}$ term represents right-handed chiral hoppings between sites in the same sublattice in the neighboring layers as shown in Fig.~\ref{fig_lattice}(b).
This term breaks inversion and mirror symmetries.
The $t_{3}$ term is a vertical interlayer hopping.
The space group of the model is $P622$.
The Bloch Hamiltonian of Eq.~(\ref{TB_Hamiltonian}) takes the following form
\begin{eqnarray}
	H_{{\bf k}} &=& d_{0} I + {\bf d}_{{\bf k}} \cdot \boldsymbol \sigma,
		\label{Bloch_Hamiltonian} \\
	d_{0} &=& 2t_{2} \cos( k_{z}c ) \sum_{i} \cos( {\bf k} \cdot {\bf b}_{i} ) \nonumber \\ 
&&\ \ \ + 2t_{3} \cos( k_{z}c ),
	\\
	d_{x} &=& t_{1}  \sum_{i} \cos( {\bf k} \cdot {\bf a}_{i} ) ,\ 	
\\ d_{y} &=& t_{1}  \sum_{i} \sin( {\bf k} \cdot {\bf a}_{i} ) ,
	\\
	d_{z} &=& -2t_{2} \sin( k_{z}c ) \sum_{i} \sin( {\bf k} \cdot {\bf b}_{i} ),
\end{eqnarray}
where ${\bf d}_{{\bf k}} = (d_{x}, d_{y}, d_{z})$, the Pauli matrices  $\sigma_i$ act on the sublattice degree of freedom, and ${\bf a}_{i}$ are vectors pointing from an A site to three neighboring B sites.
We note that the tight-binding model (\ref{Bloch_Hamiltonian}) can be mapped to the Haldane model on a honeycomb lattice \cite{PhysRevLett.61.2015} by replacing $k_z$ with a flux $\phi$.  
A similar model has been proposed in acoustic systems \cite{xiao2015synthetic}.
The orbital magnetic moment for Bloch eigenstates in this model is given by
\begin{equation}
	m_{n{\bf k}, i} = -\frac{e}{\hbar} \varepsilon_{ijl} \frac{1}{2d^{2}_{k}} {\bf d}_{{\bf k}} \cdot
		\biggl( \frac{\partial {\bf d}_{{\bf k}}}{\partial k_{j}} \times \frac{\partial {\bf d}_{\bf k}}{\partial k_{l}} \biggr),
	\label{Orbital_moment2}
\end{equation}
where $d_{k} = |{\bf d}_{{\bf k}}|$ and $\varepsilon_{ijl}$ is the Levi-Civit\'a antisymmetric tensor.
Equation (\ref{Orbital_moment2}) shows that the orbital magnetic moment is zero if any one of $d_{x}$, $d_{y}$, and $d_{z}$ is zero. 
Therefore, if $t_{1} = 0$ or $t_{2}=0$, the orbital magnetic moment vanishes at arbitrary ${\bf k}$.

\begin{figure}
\includegraphics[width=8.46cm]{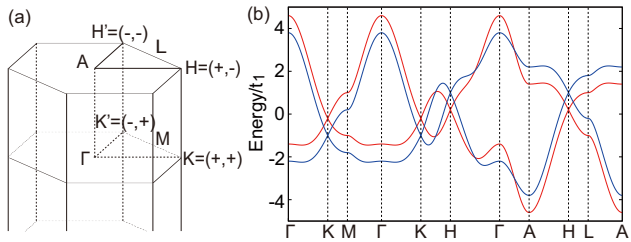}%
\caption{(a) Brillouin zone of our model with high-symmetry points.
The high-symmetry points K, K', H, H' are specified by ($\mu$, $\nu$), where 
$\mu, \nu=\pm$. 
(b) Energy bands of the Hamiltonian (\ref{Bloch_Hamiltonian}) with $t_{2} = 0.2 t_{1}$ for $t_{3} = 0.2t_{1}$ (red) and $-0.2t_{1}$ (blue).
The energy bands within $0 \leq k_{z} \leq \pi/c$ are shown.}%
\label{Fig.1}
\end{figure}

The Brillouin zone and the band structure of Eq.~(\ref{Bloch_Hamiltonian}) is shown in Fig.~\ref{Fig.1}. Our model exhibits four Weyl points, whose energies are $\varepsilon_{0} \equiv (3t_{2} - 2t_{3})$ at the H and H' points and $-\varepsilon_0=-(3t_{2} - 2t_{3})$ at the K and K' points. 
For Weyl semimetals with time-reversal symmetry but without inversion symmetry, the minimal number of Weyl points is four \cite{1367-2630-9-9-356}; therefore, this model is a minimal model for a Weyl semimetal without inversion symmetry.

\begin{figure}
\includegraphics[width=8.46cm]{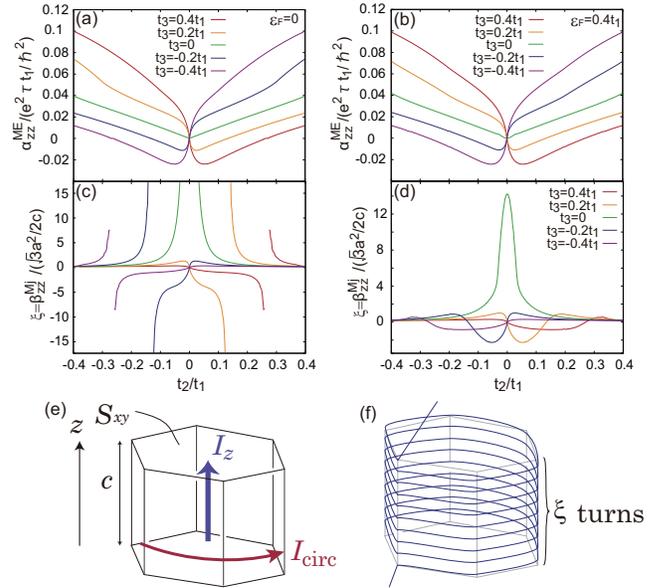}%
\caption{Numerical results of $\alpha^{\text{ME}}_{zz}$ and $\beta^{\text{Mj}}_{zz}$ from the tight-binding model (\ref{Bloch_Hamiltonian}) as a function of $t_{2}$.
(a) $\alpha^{\text{ME}}_{zz}$ at $\varepsilon_{\text{F}} = 0$.
(b) $\alpha^{\text{ME}}_{zz}$ at $\varepsilon_{\text{F}} = 0.2t_{1}$.
(c) $\beta^{\text{Mj}}_{zz}$ at $\varepsilon_{\text{F}} = 0$.
(d) $\beta^{\text{Mj}}_{zz}$ at $\varepsilon_{\text{F}} = 0.2t_{1}$. 
In (c) and (d), the vertical axes show the dimensionless 
parameter $\xi\equiv(\beta^{\text{Mj}}_{zz})/(\sqrt{3}a^2/(2c))$.
 (e) is a schematic picture of the
chiral circulating current $I_{\text{circ}}$ within a unit cell, inducing the orbital angular momentum. 
(f) shows a solenoid, which corresponds to a unit cell of a chiral crystal. We show the case with $\xi=8$, which means the solenoid with eight turns along the unit cell in the $z$ direction.}%
\label{Fig.2}
\end{figure}

The numerical results of $\alpha^{\text{ME}}$ are shown in Fig.~\ref{Fig.2}(a)(b) for $\varepsilon_{\text{F}}=0$ and $\varepsilon_{\text{F}}=0.2t_1$.
For $t_{2} = 0$, the orbital magnetization is zero due to inversion symmetry. For $t_{2} = 2t_{3}/3$, the four Weyl points are at the same energy $\varepsilon = 0$, and the orbital magnetization almost vanishes.
The numerical results of $\beta^{\text{Mj}}$ are shown in Fig.~\ref{Fig.2}(c)(d) for $\varepsilon_{\text{F}}=0$ and $\varepsilon_{\text{F}}=0.2t_1$.
For $\varepsilon_{\text{F}} = 0$ (Fig.~\ref{Fig.2}(c)), $\beta^{\text{Mj}}$ diverges at $t_{2} = 2t_{3}/3$. 
This divergence of $\beta^{\text{Mj}}$ does not mean divergence of the magnetization;
instead, the divergence of $\beta^{\text{Mj}}$ occurs because toward $t_2=2t_3/3$ the current and the orbital magnetization
in Eq.~(\ref{beta_mj}) simultaneously converge to zero, while the current density converges to zero faster than the orbital magnetization.
Meanwhile, 
as long as the current ${\bf j}$ is nonzero, $\beta^{\text{Mj}}$ never diverges and the magnetization stays finite. This behavior of $\beta^{\text{Mj}}$ is physically reasonable.
On the other hand, $\beta^{\text{Mj}}$ is finite for $\varepsilon_{\text{F}} = 0.2t_{1}$ (Fig.~\ref{Fig.2}(d)) since the current density is finite except for $t_{2} = t_{3} = 0$.

\begin{figure}
\includegraphics[width=8.46cm]{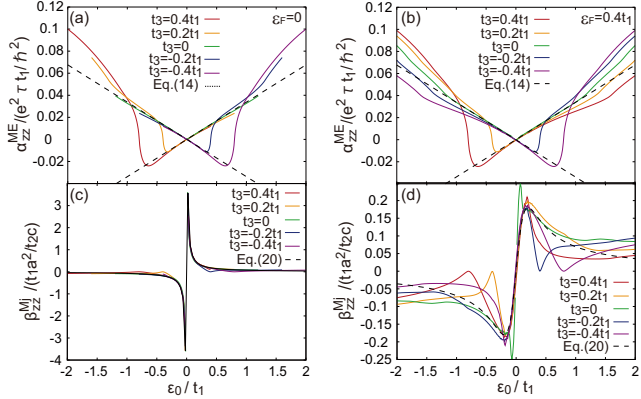}%
\caption{Numerical results of $\alpha^{\text{ME}}_{zz}$ and $\beta^{\text{Mj}}_{zz}$ as a function of $\varepsilon_{0} = 3t_{2} - 2t_{3}$. 
Comparison between the numerical result from the tight-binding model (\ref{Bloch_Hamiltonian}) and Eqs. (\ref{alpha_Weyl})(\ref{Beta_Weyl}) is shown.
(a) and (b) show $\alpha^{\text{ME}}_{zz}$, the ratio between the orbital magnetization and the electric field. The broken straight lines show the result of Eq.~(\ref{alpha_Weyl}), which is the sum of the contributions from four Weyl nodes.
$\varepsilon_{\text{F}}$ is taken as (a) $\varepsilon_{\text{F}} = 0$ and (b) $\varepsilon_{\text{F}} = 0.2t_{1}$.
(c) and (d) show $\beta^{\text{Mj}}_{zz}$, the ratio between the orbital magnetization and the current.
The broken lines show the result of Eq.~(\ref{Beta_Weyl}).
$\varepsilon_{\text{F}}$ is taken as (c) $\varepsilon_{\text{F}} = 0$ and (d) $\varepsilon_{\text{F}} = 0.2t_{1}$.
}%
\label{Fig.3}
\end{figure}

To analyze properties of $\beta^{\text{Mj}}$ in detail, we expand the Hamiltonian around one of the Weyl points up to the linear order of ${\bf k}$:
\begin{equation}
	H_{{\bf q} \mu \nu}
		= -\nu \varepsilon_{0} I - \hbar ( \mu v_{1} q_{x} \sigma_{x} + v_{2} q_{y} \sigma_{y} )
			+ \mu \nu \hbar v_{3} q_{z} \sigma_{z},
	\label{Hamiltonian_Weyl}
\end{equation}
where ${\bf q} = {\bf k} - {\bf k}_{0}$ is a displacement from the Weyl point at ${\bf k}_{0}$, $\varepsilon_{0} \equiv 3t_{2} - 2t_{3}, \hbar v_{1} = \hbar v_{2} = \frac{\sqrt{3}}{2} t_{1}a$, and $\hbar v_{3} = 3\sqrt{3} t_{2}c$.  $\mu (=\pm 1)$ and $\nu (=\pm 1)$ denote a valley degree of freedom as shown in 
Fig.~\ref{Fig.1}(a), specifying one of the Weyl points. 
We can then calculate the orbital magnetic moment around the Weyl points and it is given by
\begin{eqnarray}
	{\bf m}_{ {\bf q} \mu \nu }
		&=& - \nu \frac{e}{2} \frac{v_{1} v_{2} v_{3} }{v_{1}^{2}q_{x}^{2} + v_{2}^{2}q_{y}^{2} + v_{3}^{2}q_{z}^{2}} {\bf q},
	\label{OM_Weyl}	
\end{eqnarray}
where ${\bf q} = ( q_{x}, q_{y}, q_{z} )$. In an isotropic case  ($v_1=v_2=v_3$), this formula reduces to the result in Ref.~\citenum{PhysRevLett.116.077201}.
This result indicates that the orbital magnetic moment is enhanced around the Weyl points. Keeping only the contributions from the four Dirac cones to $\alpha^{\text{ME}}$, we obtain
\begin{equation}
	\alpha^{\text{ME}}_{ii} = \sum_{\mu \nu} \alpha^{\text{ME}}_{ii, \mu \nu}
		= \text{sgn}( v_{1} v_{2} v_{3} ) \frac{4e^{2}\tau}{3h^{2}} \varepsilon_{0},
	\label{alpha_Weyl}
\end{equation}
where $\alpha^{\text{ME}}_{ii, \mu \nu} = \text{sgn}(v_{1} v_{2} v_{3}) (e^{2}\tau/3 h^{2}) (\nu \varepsilon_{\text{F}} + \varepsilon_{0})$ is the contribution from the single Dirac cone specified by $(\mu, \nu)$.
Notably, the result for 
$\alpha^{\text{ME}}$ in an anisotropic case, Eq.~(\ref{alpha_Weyl}), is the same with 
the isotropic case obtained in Ref.~\citenum{PhysRevLett.116.077201}.
Equation (\ref{alpha_Weyl}) shows that $\alpha^{\text{ME}}$ is zero when $\varepsilon_0=0$, namely the four Weyl points are located at $\varepsilon = 0$.
While the orbital magnetic moment diverges at the Weyl point, $\alpha^{\text{ME}}$ does not diverge even when $\varepsilon_{\text{F}} = \pm \varepsilon_{0}$.
This behavior results from $q$-dependence of ${\bf m}_{{\bf q} \mu \nu}$.
The orbital magnetic moment on the Fermi surface is proportional to $q^{-1}$, and the  area of the Fermi surface is proportional to $q^{2}$.
Consequently, the Fermi surface integral of Eq.~(\ref{OM_Weyl}) per Dirac cone is roughly proportional to $q$.
We compare the numerical results from the tight-binding model and those from the effective Weyl Hamiltonian, Eq.~(\ref{Hamiltonian_Weyl}), as a function of $\varepsilon_{0} = 3t_{2} - 2t_{3}$ in Fig.~\ref{Fig.3}(a)(b).
Around $\varepsilon_{0} = 0$, the results fit well with the linear behavior, expected from Eq.~(\ref{alpha_Weyl}).
Far from $\varepsilon_{0} = 0$, the numerical result from the tight-binding model Eq.~(\ref{Bloch_Hamiltonian}) deviates from Eq.~(\ref{alpha_Weyl}) because of higher-order terms in $q$.

Next, we evaluate the tensor $\beta^{\text{Mj}}$ describing the ratio between the current and the orbital magnetization, defined in Eq.~\ref{beta_mj}. 
By using $j_{i} = \sigma_{ij} E_{j}$, where $\sigma_{ij}$ is the conductivity, $\alpha^{\text{ME}}$ can be written as $\alpha_{ij}^{\text{ME}} = \beta_{ik}^{\text{Mj}} \sigma_{kj}$.
The conductivity is given by the Fermi-surface integral,
\begin{equation}
\sigma_{ij} = -e\tau \sum_{n} \int_{\text{BZ}} \frac{d{\bf k}}{(2\pi)^{3}}
		\frac{ df } { d\varepsilon } \Big|_{\varepsilon = \varepsilon_{n{\bf k}}}
		v_{n{\bf k}, i} v_{n{\bf k}, j},
\end{equation}
where the Boltzmann transport theory with relaxation-time approximation is adopted. 
The conductivity for the effective Hamiltonian Eq.~(\ref{Hamiltonian_Weyl}) is calculated as
\begin{align}
&\sigma_{xx} = \sigma_{yy} = \frac{8}{3} \frac{e^{2} \tau }{h^{2}} \frac{\varepsilon_{\text{F}}^{2} + \varepsilon_{0}^{2} }{\hbar | v_{3} |},
\label{Conductivity_Weyl}\\
&\sigma_{zz} = \frac{8}{3} \frac{e^{2} \tau }{h^{2}} \frac{|v_{3}| ( \varepsilon_{\text{F}}^{2} + \varepsilon_{0}^{2} )}{\hbar |v_{1}| |v_{2}| }.
\label{Conductivity_Weyl2}
\end{align}
With Eqs.~(\ref{alpha_Weyl}),  (\ref{Conductivity_Weyl}), (\ref{Conductivity_Weyl2}), and $\alpha^{\text{ME}}_{ij} = \beta^{\text{Mj}}_{ik} \sigma_{kj}$, we obtain
\begin{align}
&\beta^{\text{Mj}}_{xx} 
		=  \frac{\hbar v_{2} \hbar v_{3}}{2\hbar v_{1}}
			\frac{ \varepsilon_{0} }{ \varepsilon_{\text{F}}^{2} + \varepsilon_{0}^{2} }
		= \frac{3\sqrt{3}}{2} \frac{ t_{2}( 3t_{2} - 2t_{3} ) }{ \varepsilon_{\text{F}}^{2} + ( 3t_{2} - 2t_{3} )^{2} } c,
		\\
	&\beta^{\text{Mj}}_{yy}
		=  \frac{\hbar v_{3} \hbar v_{1}}{2\hbar v_{2}}
			\frac{ \varepsilon_{0} }{ \varepsilon_{\text{F}}^{2} + \varepsilon_{0}^{2} }
		= \frac{3\sqrt{3}}{2} \frac{ t_{2}( 3t_{2} - 2t_{3} ) }{ \varepsilon_{\text{F}}^{2} + ( 3t_{2} - 2t_{3} )^{2} } c,
		\\
	&\beta^{\text{Mj}}_{zz}
		=  \frac{\hbar v_{1} \hbar v_{2}}{2\hbar v_{3}}
			\frac{ \varepsilon_{0} }{ \varepsilon_{\text{F}}^{2} + \varepsilon_{0}^{2} }
		= \frac{\sqrt{3}}{24} \frac{t_{1}^{2}a^{2}}{t_{2}c} \frac{ 3t_{2} - 2t_{3} }{ \varepsilon_{\text{F}}^{2} + ( 3t_{2} - 2t_{3} )^{2} }.
	\label{Beta_Weyl}
\end{align}
When $\varepsilon_{\text{F}} \neq 0$, $\beta^{\text{Mj}}_{zz}$ is zero at $3 t_{2} = 2t_{3}$ because of $\alpha^{\text{ME}}_{zz} = 0$ and $\sigma_{zz} \neq 0$. Meanwhile, when $\varepsilon_{\text{F}} = 0$, both $\alpha^{\text{ME}}_{zz}$ and $\sigma_{zz}$ are zero, which leads to a divergent behavior of $\beta^{\text{Mj}}_{zz}$ at $3t_{2} = 2t_{3}$. 
This divergence arises because $\sigma_{zz}$ converges to zero faster than $\alpha^{\text{ME}}_{zz}$, 
as $\varepsilon_0$ approaches zero.
In other words, as previously mentioned, for the fixed electric field and at $\varepsilon_{\text{F}}=0$, both the current 
${\bf j}$ and the magnetization ${\bf M}$ go to zero as $\varepsilon_0$ approaches 
zero, but ${\bf j}$ approaches zero faster than ${\bf M}$, leading to the divergence
of  $\beta^{\text{Mj}}_{zz}$ from Eq.~(\ref{beta_mj}).
We compare the results from the  tight-binding model with those from the Weyl Hamiltonian, Eq.~(\ref{Hamiltonian_Weyl}), in Fig.~\ref{Fig.3}(c) and (d). They agree well with each other  around $\varepsilon_{0} = 0$. When $\varepsilon_{0}$ is far from zero, the approximation as a Weyl Hamiltonian is no longer valid due to higher-order terms in $q$.

Current-induced orbital magnetization, i.e. the orbital Edelstein effect, was previously studied in tellurium \cite{ivchenko1978new,vorob1979optical,Shalygin2012} and in zeolite-templated carbon \cite{PhysRevB.86.125207}. However, the physical mechanism of our study totally differs from that of the previous studies. The orbital magnetization studied previously \cite{ivchenko1978new,vorob1979optical,Shalygin2012,PhysRevB.86.125207} originates from  an intracelluler or intrasite circulation current from atomic orbitals of each atom.
In our theory \cite{PhysRevLett.95.137204,PhysRevLett.95.137205,PhysRevB.74.024408,PhysRevLett.99.197202,RevModPhys.82.1959,0953-8984-22-12-123201}, the orbital magnetic moment ${\bf m}_{n{\bf k}}$ originates from intercellular or intersite circulation current \cite{Yafet1963,PhysRevB.77.235406}.
In our model Eq.~(\ref{Bloch_Hamiltonian}), no intrasite orbital magnetization is induced because we assume that the localized basis of the model Eq.~(\ref{Bloch_Hamiltonian}) has no orbital angular momentum.
The intrasite and intersite orbital Edelstein effects do not need the spin-orbit interaction.
Furthermore, the intersite orbital Edelstein effect does not need the orbital angular momentum of the parent atomic orbital.
The intersite orbital Edelstein effect requires gyrotropic lattice structure, in the same way as the intrasite orbital Edelstein effect and the spin Edelstein effect. Meanwhile,  
geometrical structure of crystals is expected to be significant in the intersite orbital one, 
compared with the other two effects. 
In this sense, this orbital magnetization might have some similarity with 
valley physics \cite{PhysRevLett.99.236809,PhysRevLett.108.196802}, since they both come from lattice properties. 
The orbital Edelstein effect can be measured experimentally, but 
in the measurement of magnetization, one should separate the spin and orbital parts. Here, intrasite and intersite orbital Edelstein effects are 
always mixed, because the distinction between them lies in the description of eigenstates 
in terms of 
atomic orbitals. Such a description becomes suitable in the limit of large interatomic separations, but 
is not accurate in crystals, leading to mixing between the intersite and intrasite orbital
Edelstein effects.
Recently, an NMR shift in tellurium with a current has been measured, and the resulting shift 
is proportional to the current \cite{Furukawa17}. Both the spin and the orbital parts
may contribute to the shift, and their separation requires comparison with numerical calculations.

The Bohr magneton, which is a fundamental unit of the magnetic moment of an electron, does not appear in Eq.~(\ref{Orbital_moment}).
In our model, the orbital magnetic moment Eq.~(\ref{Orbital_moment}) is measured as a unit of $eta^{2}/2\hbar$ instead of the Bohr magneton, where $t$ is an energy scale of the Hamiltonian such as the nearest-neighbor hopping energy and $a$ is the lattice constant. 
At the bottom of the conduction band or at the top of the valence band, the inverse effective mass $1/m^{*}$ is of the order of $ta^{2}/\hbar^{2}$.
Therefore the unit $eta^{2}/2\hbar = (e\hbar / 2(\hbar^{2}/ta^{2}))$ can be  interpreted as a magnetic moment generated by an electron with a mass $\hbar^{2}/ta^{2} \sim m^{*}$.
At the K point of graphene with a staggered potential, the orbital magnetic moment agrees with the effective Bohr magneton $\mu_{\text{B}}^{*} = e\hbar / 2m^{*}$ \cite{PhysRevLett.99.236809,PhysRevB.77.235406,RevModPhys.82.1959}.
The unit $eta^{2}/2\hbar$ can be also expressed as $eta^{2}/2\hbar = IS$, where $I = e(ta/\hbar) / 2\pi a$ is a circulating current with a radius $a$ and velocity $ta/\hbar$, and $S = \pi a^{2}$.

Here, we compare the size of the orbital Edelstein effect in our model with that of a solenoid in classical electrodynamics.
First, we evaluate $\beta^{\text{Mj}}$, by modelling a chiral crystal in the following way.
We set the axis of the crystal be along the $z$-direction.
The unit cell has a length $c$ along the $z$ axis and an area $S_{xy}$ along the $xy$ plane.
We shall decompose the current flowing in the unit cell into two; one is a current $I$ along the $z$ axis, and the other is a circulating current $I_{\text{circ}}$ around the unit cell within the $xy$ plane, as shown in Fig.~\ref{Fig.2}(e).
From classical electrodynamics, the circulating current induces an orbital magnetic moment $m_{z} \simeq I_{\text{circ}}S_{xy}$, and  the resulting orbital magnetization $M_{z}$ is given by $M_{z}  = m_{z} / S_{xy}c \simeq I_{\text{circ}} / c$. On the other hand, the electric current density $j_{z}$ is evaluated as $j_{z} \simeq I/S_{xy}$.
Therefore, the tensor $\beta^{\text{Mj}}_{zz}$ is represented as $\beta^{\text{Mj}}_{zz} \simeq \xi (S_{xy}/c)$, where $\xi = I_{\text{circ}} / I$. This dimensionless factor $\xi$ indicates 
the ratio between two components of the current, shown in Fig.~\ref{Fig.2}(e).

One can draw an analogy with a classical solenoid, and evaluate efficiency of the orbital Edelstein effect in a chiral crystal. 
In a solenoid with the number of turns per unit length $n$, an electric current $I$ induces a magnetic field $H = n I$; therefore, a magnetic moment per volume is given by $M=nI$.
By comparing this formula with the corresponding formula of $M_z$ 
for the orbital Edelstein effect, 
we obtain $\xi =nc$; namely, the dimensionless factor $\xi$ indicates the number of turns within the unit cell when the crystal is regarded as a classical solenoid (see Fig.~\ref{Fig.2}(f)).
Thus, this dimensionless factor $\xi$ represents how ``chiral" the crystal is, and classically it is of the order of unity. Remarkably, as seen in Fig.~\ref{Fig.2}(c), the value of $\xi$ can be much larger than that expected from a 
geometrical structure of the crystal. For example, in the crystal structure shown in Fig.~\ref{fig_lattice}(b), one may naively expect this dimensionless factor $\xi$ to be maximally $1/3$, but it is not true.
In our model with $\varepsilon_{\text{F}} = 0$, $\xi$ diverges as $\varepsilon_0$ approaches zero (Fig.~\ref{Fig.2}(c)). Namely, $\xi$ becomes large for a Weyl semimetal, with all the Weyl nodes close to the Fermi energy.
Thus,  counterintuitively, the crystal structure in Fig.~\ref{fig_lattice}(b) works as a solenoid with 
many turns within the unit cell (see Fig.~\ref{Fig.2}(f) for $\xi=8$ as an example) . 

How a chiral crystal compares with a classical solenoid from the viewpoint of current-induced magnetization depends on the size of the system. 
In nanoscale, the size of the current-induced magnetization for a chiral crystal can 
exceed that of a classical solenoid, as $\xi$ can exceed unity. As the scale becomes larger, 
a chiral crystal becomes less favorable for inducing large magnetization, compared with a classical solenoid; it is  
because for a fixed amount of the current, the current density inversely scales with the area of the system within the $xy$ plane, and the magnetization scales similarly, while for 
a classical solenoid, the area within the $xy$ plane does not affect the strength of the generated magnetic field. Thus to make a nanoscale ``solenoid'', the orbital Edelstein effect in a chiral crystal 
may be useful compared with a classical solenoid. For this purpose, a Weyl semimetal with 
all the Weyl nodes located near the Fermi energy is ideal, and such a Weyl semimetal can be found close to a phase transition to an insulator phase, because the phase transition is
accompanied by creations of Weyl nodes. 

In conclusion, we have defined the tensor $\beta^{\text{Mj}}$ to quantify the orbital Edelstein effect, and clarified the analogy between $\beta^{\text{Mj}}$ and a solenoid. The tensor $\beta^{\text{Mj}}$ is characterized by the dimensionless factor $\xi$ and lattice constants.
We show that $\beta^{\text{Mj}}$ and $\xi$ can be enhanced by designing band structure, in particular, for a Weyl semimetal with all the Weyl nodes close to the Fermi energy.

\section*{Author Information}

{\bf Corresponding Authors}\\
$^*$E-mail: murakami@stat.phys.titech.ac.jp.

\noindent
{\bf Notes}\\
The authors declare no competing financial interest.

\begin{acknowledgement}
T. Yoda is a JSPS Research Fellow.
This work was supported by JSPS KAKENHI Grant No. JP16J07354 and 
by Grants-in-Aid for Scientific Research on Innovative Areas ``Topological Materials Science'' (KAKENHI Grant No. JP16H00988) 
and ``Nano Spin Conversion Science'' (Grant No. JP17H05179 and 26103006), by MEXT Elements Strategy Initiative to Form Core Research Center (TIES), and
by  JSPS KAKENHI Grant Number JP16K13834.
\end{acknowledgement}

\bibliography{Refs}

\end{document}